\documentclass[sn-vancouver,Numbered]{sn-jnl}
\usepackage{lmodern}
\usepackage{graphicx}%
\usepackage{multirow}%
\usepackage{amsmath,amssymb,amsfonts}%
\usepackage{amsthm}%
\usepackage{mathrsfs}%
\usepackage[title]{appendix}%
\usepackage{xcolor}%
\usepackage{textcomp}%
\usepackage{manyfoot}%
\usepackage{booktabs}%
\usepackage{algorithm}%
\usepackage{algorithmicx}%
\usepackage{algpseudocode}%
\usepackage{physics}%
\usepackage{mathtools}%
\usepackage{tabularx}%
\usepackage[T1]{fontenc}
\DeclareUnicodeCharacter{2212}{-}
\usepackage{subfig}

\begin{document}

\title[Article Title]{EHCube4P: Learning Epistatic Patterns Through Hypercube Graph Convolution Neural Network for Protein Fitness Function Estimation}
\author[1,2]{\fnm{Muhammad}\sur{Daud}}
\email{m.daud.sns@gmail.com}
\author[2]{\fnm{Philippe}\sur{Charton}}
\email{philippe.charton@univ-reunion.fr}
\author[3]{\fnm{Cedric}\sur{Damour}}
\email{cedric.damour@univ-reunion.fr}
\author[1]{\fnm{Jingbo}\sur{Wang}}
\email{jingbo.wang@uwa.edu.au}
\author*[2,4]{\fnm{Frederic}\sur{Cadet}}
\email{frederic.cadet.run@gmail.com}

\affil[1]{\orgdiv{Centre for Quantum Information, 	
Simulation and Algorithms}, \orgname{School of Physics, Mathematics and Computing, The University of Western Australia}, \orgaddress{\city{Perth} \state{WA} \country{Australia}}}
\affil[2]{\orgname{University of Paris City \& University of Reunion}, \orgdiv{BIGR, Inserm, UMR S1134}, \orgaddress{\city{Paris}, \country{France}}}
\affil[3]{\orgdiv{EnergyLab, EA 4079}, \orgname{University of Reunion}, \orgaddress{\city{Saint-Denis}, \country{France}}}
\affil[4]{\orgdiv{PEACCEL}, \orgname{Artificial Intelligence Department, AI for Biologics}, \orgaddress{\city{Paris}, \country{France}}}

\abstract{Understanding the relationship between protein sequences and their functions is fundamental to protein engineering, but this task is hindered by the combinatorially vast sequence space and the experimental noise inherent in fitness measurements. In this study, we present a novel framework that models the sequence landscape as a hypercube $H(k,2)$ and integrates wavelet-based signal denoising with a graph convolutional neural network (GCN) to predict protein fitness across rugged fitness landscapes. Using a dataset of 419 experimentally measured mutant sequences of the Tobacco 5-Epi-Aristolochene Synthase (TEAS) enzyme, we preprocess the fitness signals using a 1-D discrete wavelet transform with a Daubechies-3 basis to suppress experimental noise while preserving local epistatic patterns. Our model comprises two GCN layers, allowing for beyond pairwise aggregation, followed by a multi-layer perceptron (MLP). We show that our approach, EHCube4P, generalizes well across different enzyme activity datasets and effectively captures higher-order mutational interactions. Performance varies with the ruggedness of the fitness landscape, with smoother signals yielding higher test set $r^2$ scores. These results demonstrate that combining wavelet preprocessing with graph-based deep learning enhances the robustness and generalization of fitness prediction, particularly for sparse and noisy biological datasets. The approach provides a scalable and interpretable framework for protein fitness estimation applicable to a broad range of combinatorial biological systems.}
\maketitle
\section{Introduction}\label{sec1}
Proteins play a vital role in sustaining life. The identification of non-linear effects of protein function due to interactions among amino acids, epistasis, and their molecular dynamics is crucial to understanding protein function. The protein fitness landscape allows for many possible evolutionary pathways, yet most of these remain largely unexplored in laboratory evolution studies~\cite{Acevedo-Rocha2021}. Scientists have developed various approaches to investigate protein sequences, their structures, and functions~\cite{CRF2022}. One of the most significant advances in this area is \textit{gene editing}, which enables the manipulation of protein properties such as enzyme activity, fluorescence, and binding affinity. Protein engineering, in particular, focuses on modifying amino acid sequences to alter the overall function and characteristics of proteins~\cite{DLandPE}. Directed evolution and rational design offer valuable insights into the effects of amino acid interaction on protein design and their function~\cite{Cadet2022,Wenran2024}. The relationship between a protein’s sequence and its function is often conceptualized as a “fitness landscape”, which maps every possible amino acid variation to a corresponding phenotypic value, often referred to as fitness~\cite{Wittmund2022}. The central goal of protein engineering is to explore this landscape in search of high-fitness sequences~\cite{Romero2009}.

Searching within the fitness landscape is an extremely challenging task. The reason is primarily due to the following factors: (1) exponentially large sequence space, (2) fragmentary fitness data, and (3) lack of biological information about the mutant sequence. Constructing the full sequence space is virtually impossible, given the prohibitive time and financial costs involved.~\cite{MSACHOWDHURY2017419}. One approach to exploring sequence variation is the single-site saturation method, where each residue position is systematically replaced with alternative amino acids. However, this leads to a rapid combinatorial explosion of the sequence space: with $k$ mutable positions and $n$ possible amino acids at each site, the total number of variants grows exponentially to $n^k$. In practice, due to high costs and the absence of efficient high-throughput screening methods, the vast majority of these variants remain unexplored in the wet lab, leaving us with a fragmentary description of the sequence-fitness relationship~\cite{VisualingFitnessSpace}. 

We can model these interactions as a Hamming graph $H(k,n)$. A special case, when only 2 amino acids are considered at each residue position, makes a hypercube $H(k,2)$.
Although the relation between sequence and fitness is an open question, it is widely accepted that epistasis, non-linear interactions of amino acids at residue positions, play an important role in observing the fitness of the entire sequence space~\cite{Poelwijk2019}.

Two of the most common approaches to protein engineering are laboratory-based directed evolution~\cite{FCadet2018,MSACHOWDHURY2017419} and computational physics-based rational design~\cite{LEVY201755}. Machine learning based on directed evolution uses the fitness constraints from naturally occurring proteins called evolutionary data. Then a probability model such as the Hidden Markov Model is used to predict the relative phenotype value of the mutant sequence~\cite{Wenran2024}. Such models, though effective, require large amounts of protein evolutionary data about the specific property of interest~\cite{Biswas2021}. The other approach of the supervised regression model $f(\cdot)$, where some input feature $x$ are used to approximate label $y$ by minimizing some objective function e.g. $||f(x)-y||^2$. These models use mutant sequences coupled with experimentally determined assay labels for the property of interest~\cite{Hsu2022}. Such models, though useful, are limited to a few residue positions. One of the simplest models is linear regression, which uses amino acid features to model pairwise interactions, to a good approximation. Linear models do not provide a complete picture of the fitness landscape, as they do not take into account higher-order interactions~\cite{HOE&PP}. 

Recent advances in deep learning and graph representation learning have shown that structured data, when represented as a graph, can be used to model complex connections between features~\cite{sanchez-lengeling2021a,GNN_IN_BI}. A pioneering work~\cite{WANG2023107952} has shown that a graph neural network augmented with a multi-layer perceptron can be used to obtain improved generalization and cater to the high-throughput requirements of the industry. Their model utilizes evolutionary data to extract the molecular structure of the mutant amino acid to learn the complex relation between the molecular structure of the wild-type protein and fitness. This model is only limited to a few residue positions $(\sim 4)$ because the parameter space becomes exponentially large for a large number of residues.

This article presents a novel deep learning approach, EHCube4P, to learn epistatic patterns of the sequence space to make sound predictions about the protein fitness function. The EHCube4P introduces a novel method of visualizing the fitness landscape as a Hamming graph, and utilizing wavelet-based denoising of experimental data. We use two graph convolution layers, GCN-1 and GCN-2, the first layer allows aggregation from adjacent mutant vertices, and the second layer allows information aggregation from beyond adjacent mutants. The two GCN layers plus a Multi-Layered Perceptron~(MLP) allow us to explore beyond pairwise interactions of amino acids within the sequence space to approximate non-linear behavior of phenotype values for each sequence. Our novel approach shows that denoising the experimental fitness data causes the machine learning models to show improved generalization and avoid overfitting. The aggregation of information from neighboring mutant sequences helps the model to learn complex interaction patterns of the fitness space. We present our claims by predicting fitness values through modeling of beyond pairwise interactions in one of the standard datasets that has a rugged fitness landscape, referring to epistasis~\cite{O'Maille2008}. The structure of the article is as follows: firstly, Sec.~\ref{sec1} provides an introduction to the problem and highlights the novelty of the study. Sec.~\ref{sec2} provides details about the methods we used during the course of our study. Sec.~\ref{Exp Data} provides insights about the choice of dataset, Sec.~\ref{Data denoising} presents our novel approach to remove experimental noise, Sec.~\ref{Graph model} presents the new way of modeling sequence-fitness landscape that helps deep learning model explained in Sec.~\ref{GCN} to learn complex patterns within the protein sequence. Lastly, we present our results in Sec.~\ref{Results}, and conclude our findings in Sec.~\ref {Conc}.
\section{Methods}\label{sec2}
\subsection{Experimental Data and Epistasis}\label{Exp Data}
One of the first datasets for understanding context dependence of mutational effect - Epistasis, is first reported in ref.~\cite{O'Maille2008}. It is taken as a standard in the field for the characterization of the catalytic landscape of sesquiterpene synthase, where mutational changes alter the biosynthetic properties of the enzyme. They created the complete mutational library for nine naturally occurring mutations interconverting the two orthologous sesquiterpene synthase, tobacco 5-epi-aristolochene synthase (TEAS) and henbane premnaspirodiene synthase (HPS). Product spectra through gas chromatogram mass spectrometry were reported for $419$ out of $2^9=512$ mutants, for $5$-Epi-Aristolochene ($5$EA), $4$-Epi-Eremophilene($4$EE), premnaspirodiene (PSD), and other minor products(MP). The product spectra revealed a rugged landscape, which is associated with the presence of epistasis. Complex functional variability in product spectra is not readily predictable based on evolutionary analysis~\cite{Koo2016}. Their study speculates that the $4$-EE might be the dominant product of the 9-Mutation library due to its common TEAS-HPS ancestry. We aim to model the reported data, with the understanding that the rugged landscape or the strong local variation reported in the experimental fitness values are due to epistatic interactions caused by mutations in the wild-type sequence.

\subsection{\label{Data denoising} Data preparation and Denoising Technique}

Modeling the mutational interactions of amino acids is a challenging task, mainly due to experimental noise. Avoiding global experimental noise from the data is almost inevitable because amino acid interactions cause strong variations in the function value of the property of interest(fitness). Thus, the global non-linear behavior of the experimental fitness values can arise due to epistasis or noise~\cite{Poelwijkbook}. We can model the experimental fitness as:
\begin{equation}\label{fiteq}
    y(\bar{x})=\mathrm{f}(\bar{x})+\epsilon,
\end{equation}
where $y(\bar{x})$ is the experimental fitness from assay, $\mathrm{f}(\bar{x})$ is the unknown function that maps the protein sequence to a noiseless phenotype value i.e $f:\{0,1\}^{\otimes k}\longrightarrow \mathbb{R}$, and $\epsilon$ is the random global white noise. As the available experimental fitness is rarely complete, it makes it difficult to remove the noise in an unbiased manner. We utilize a well-studied technique from signal processing for denoising unknown signals through the discrete wavelet transform~\cite{SCHNEIDER2006426}. Before that we introduce two fundamental definitions important for DWT application. 
\paragraph{Definition 1: fitness map as signal}
Let $\mathcal{X}\subseteq\{0,1\}^{\otimes k}$ be the space of all binary sequences of length $k$. Then $f:\mathcal{X}\longrightarrow\mathbb{R}$ denote a function assigning a real-valued scalar to a property of interest to each sequence. The domain $\mathcal{X}$ is finite, discrete, and can be logically structured over a graph.

\paragraph{Definition 2: Gray Code Ordering}
Let $\mathcal{X} = \{0,1\}^k$ represent the binary sequence space of $k$ mutational sites, forming a $k$-dimensional hypercube.
Then, a \emph{Gray Code Ordering} is an arrangement or ordering of binary sequences such that each consecutive pair differs by exactly one bit. It provides a biologically meaningful linearization of the hypercube-structured sequence space, reflecting a single point mutation, and preserving locality.
\newline

A standard denoising procedure includes three steps: (1) decomposition of signal via Discrete Wavelet Transform~(DWT) into approximate and detail coefficients, (2) thresholding of detail coefficients, (3) reconstruction of signal from thresholded coefficients~\cite{mallat1999wavelet}. We apply this method by considering the experimental fitness values as $M$ number of samples of an underlying unknown protein fitness function $y(\bar{x})$, which maps the sequence $\bar{x}$ to a unique fitness value. Even though the samples might not provide the exact behavior, a 1-D DWT tries to approximate the sampled function by scaling and shifting a wavelet basis function $\psi(x)$,
\begin{equation}
    y(\bar{x}_m)=\sum_ia_i\psi_i(\bar{x}_m)\quad\forall\quad m\in [0,M).
\end{equation}
Here, the wavelet coefficients $a_i$ are filtered into two categories, approximate $\bar{c}_a$ and detail $\bar{c}_d$ coefficients. We refer the reader to ref.~\cite{WTR,MOULIN2005347,mallat1999wavelet} for an in-depth overview. The approximate coefficients, which record the underlying behavior of the function, are kept intact. But the detail coefficients that store information about sudden variations in the signal are thresholded using the universal threshold~\cite{WT_UT} defined as:
\begin{equation}
    \lambda=\sigma\sqrt{2\log M}
\end{equation}
where $\sigma=\frac{\mathrm{Median}(|\bar{c}_d|)}{0.6745}$, because for a standard normal distribution, $\text{median absolute deviation}\approx 0.6745\sigma$. We use the soft thresholding technique~\cite{donoho1994ideal}, which shrinks all coefficients smaller than $\lambda$ to zero and reduces all larger coefficients by $\lambda$. i.e.
\begin{equation}
    c^\prime_d=\begin{cases}
        \mathrm{sign}(c)(|c|-\lambda)\quad\mathrm{if}\quad |c|\ge\lambda\\
        0 \qquad\qquad\quad\qquad\mathrm{if}\quad |c|<\lambda
    \end{cases}.
\end{equation}
We can recover the denoised signal with the help of an inverse wavelet transform on these thresholded coefficients.
\subsection{\label{Graph model}Modeling Fitness Landscape as Graph}
Proteins are long chains of amino acids that fold into a complex 3-D structure to perform some function. Mutations can occur naturally or can be introduced through guided laboratory techniques~\cite{Chatterjee2017-up}. Due to the nature of the problem, we only model specific mutation sites called residue positions,$k$. Ideally, all $20$ naturally occurring amino acids can take up all available residue positions, making the sequence space exponentially large $\sim 20^k$. This makes the problem of searching the sequence space for high fitness sequences very costly. We derive inspiration from point mutations~\cite{Brown2002-uv}, which allows us to introduce iterative changes in the sequence to get a completely mutated sequence. This can be modeled as a Hamming graph $H(k,n)$, where $n$ is the number of unique amino acids at each position and $k$ is the total number of residue positions. Graph structure allows each vertex, which is a sequence, to be one change away from its neighbouring sequences, and each link represents an interaction between the amino acid residues.
This approach can model more than two amino acid residues and any number of residue positions, providing an efficient way to conduct evolutionary studies. We use an $H(9,2)$ hypercube graph to model the enzyme activity Tobacco 5-Epi-Aristolochlene synthase for the production of different molecules, see Supp.~Fig.~S1.

\vspace{-0.6cm}
\begin{figure}[!ht]
    \centering \includegraphics[width=0.6\linewidth]{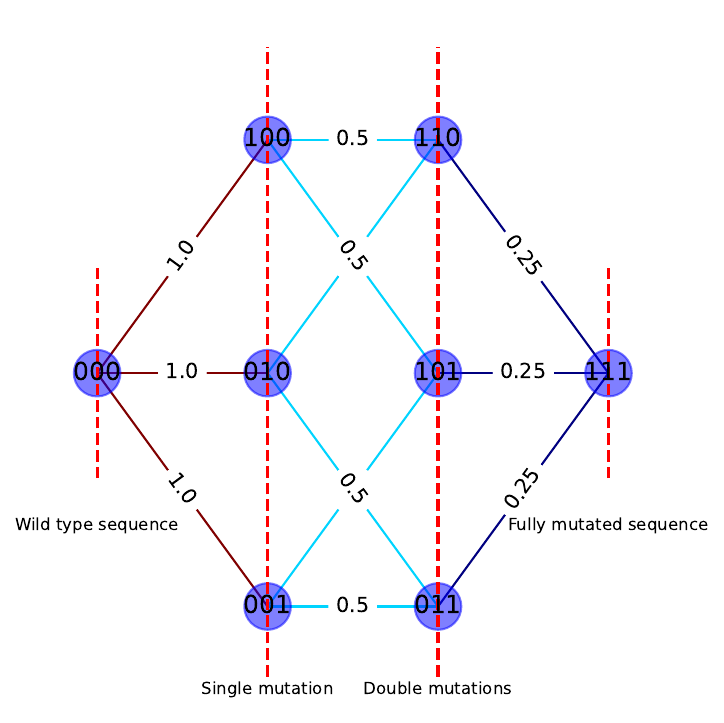}
    \caption{A representation of sequence space for three residue positions, each position taking up only two amino acids(AA), wild-type AA and mutant AA. The resulting space is a special case of Hamming graph, $H(k,2)$, a hypercube.}
    \label{HG example}
\end{figure}

As it is highly inefficient to get enough information about the entire $20^k$ sequences, the majority of the space remains unexplored for large $k$. This can be tackled if we only allow the residue position to take up two amino acids, a wild-type AA or a mutant AA, represented by $0$ or $1$, respectively. Such a sequence space $\sim 2^k$ can be mapped to a hypercube graph $H(k,2)$, a special case of the Hamming graph, an example is shown in Fig.~\ref{HG example}. A recursive definition of the adjacency matrix of the hypercube is given as:
\begin{equation}
    A_k=
    \begin{cases}
        \mathbb{I}_2\otimes A_{k-1}+A_1\otimes\mathbb{I}_{2^{k-1}} \quad\text{if}\quad k>1 \\
        \begin{bmatrix}
            0&1\\1&0
        \end{bmatrix}
        \quad \text{if}\quad k=1
    \end{cases}.
\end{equation}
As reported by ref.~\cite{Poelwijk2016, Poelwijk2019}, it is difficult to capture epistasis at higher orders of mutation due to the global noise. We assign weights to the edges depending on the order of interactions, i.e. $\frac{1}{2^{i\odot j}}$, where $i\odot j$ is the bit-wise dot product between the two vertices connected via the edge. Uneven edge weight leads to an undesirable matrix spectrum, which can be circumvented by normalizing the Laplacian matrix: $L_{norm}=D^\frac{-1}{2}LD^\frac{-1}{2}=I-D^\frac{-1}{2}AD^\frac{-1}{2}$, where Laplacian matrix is defined as $L=D-A$ and $D$ is the degree matrix.
\subsection{\label{GCN}Graph Convolution Network}

\begin{figure}[!ht]
    \centering
    \includegraphics[width=\textwidth]{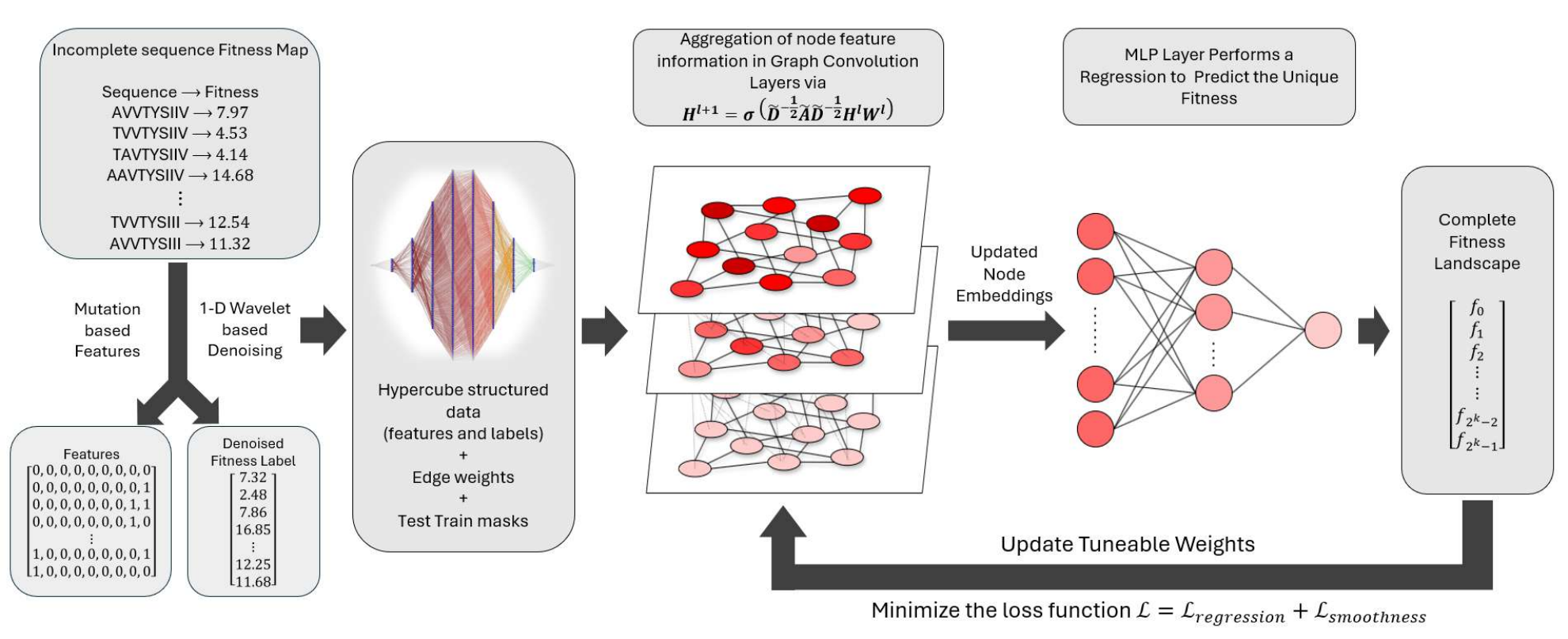}
    \caption{\protect A schematic diagram for the protein fitness estimation GCN. Aggregation of information in GCN layer highlights the highly functional nodes(mutants) in the graph(sequence space). The updated node representations are passed as input to the MLP which predict fitness values.}
    \label{GCN model}
\end{figure}

Graph convolution networks (GCN) are a powerful deep learning tool for performing various graph-structured data tasks. We use GCN for a node-level regression task. Our graph model treats each mutant sequence as a node, and we use a GCN and an MLP to predict the continuous fitness value, as shown in Fig.~\ref{GCN model}.

We formulate the mutant feature vector $\bar{X}_i$ where $i\in \{0,\dots,N-1\}$ where $N=2^k$, to be a binary vector $\bar{X}_i=[x_0,x_1,x_2,\dots,x_{k-1}]$ where $x_j \in \{0,1\}$. Thus, the input feature matrix for GCN is $\mathbf{X}\in \mathbb{R}^{N\times k}$. Graph structure explains the interactions as weighted edges and connectivity, stored in the adjacency matrix $\mathbf{A}\in \mathbb{R}^{N\times N}$ and the degree matrix $\mathbf{D}\in \mathbb{R}^{N\times N}$. Graph representation is learned via aggregating mutant feature information from the neighboring vertices~\cite{gori2005new} via:
\begin{equation}
    \bar{h}_i^{l+1}=\sigma\Big(a\sum_{j\in\mathcal{N_i}}\frac{\bar{h}^l_j}{\sqrt{d_j}\sqrt{d_i}}+b\bar{h}_i^l\Big),
    \label{agg_eq}
\end{equation}
where $\sigma$ is a non-linear activation function, $\bar{h}_i^l$ is the $l^{th}$ latent feature vector for $i^{th}$ vertex with $\bar{h}_i^0=\bar{X}_i$,  and $\mathcal{N}_i$ is the set of neighboring vertices for vertex $i$, and $a,b$ are tunable weights. Eq.~\ref{agg_eq} can be expressed in terms of matrix multiplication, which is more suited for graph-structured data. Eq.~\ref{GCNeq} provides the general view of the aggregation of vertex information\cite{kipf2017semisupervised},
\begin{equation}
    \mathbf{H}^{l+1}= \sigma \Big( \tilde{\mathbf{D}}^{-\frac{1}{2}}\tilde{\mathbf{A}}\tilde{\mathbf{D}}^{-\frac{1}{2}}\mathbf{H}^l\mathbf{W}^l\Big),
    \label{GCNeq}
\end{equation}
where $\tilde{\mathbf{A}}=\mathbf{I}+\mathbf{A}$ is equivalent to adding self-loops to the underlying graphs. $\tilde{\mathbf{D}}$ is degree matrix of $\tilde{\mathbf{A}}$, $\mathbf{H}^l\in\mathbb{R}^{N\times \dim(l)}$ is the latent feature matrix for $l^{th}$ convolution layer, and we use $\tanh$ as non-linear activation function $\sigma$. $\mathbf{W}^l$ is the learnable weight matrix of dimensions $\dim(l)\times \dim(l+1)$.

Given that we only have experimental fitness labels for $m<N$ mutants. We aim to learn the interaction pattern from the labeled mutants and predict the $N-m$ missing mutant fitness. Our downstream task becomes a semi-supervised regression task, where unlabeled mutants participate in vertex aggregation but do not participate in training. The downstream task is through a multi-layered perceptron layer, which takes in the updated node embeddings as its input and outputs a continuous fitness value. Thus, the model can be represented as:
\begin{equation}
    \bar{y}=\mathbf{MLP\big(GCN(A,(X,W))\big)}.
    \label{CompGCNeq}
\end{equation}
Here, the $\mathbf{MLP}$ represents a sequential two-layered linear neural network with $\mathrm{ReLU}$ as non-linear activation function, and the $\bar{y}$ is the estimated fitness values. The training part only uses the unmasked/labeled vertices. We set their fitness as the target we want to achieve, and the goal is to minimize the loss function for labeled data~\cite{ma2019flexiblegenerativeframeworkgra}. The loss function is defined as:
\begin{equation*}
    \mathcal{L}=0.7*\mathcal{L}_\mathrm{regression}+0.3*\mathcal{L}_\mathrm{smoothness},
\end{equation*}
\begin{equation}
    \mathcal{L}_{\mathrm{regression}}= \frac{1}{|m|}\sum_{i=0}^{m}||y_{i_{\mathrm{pred}}}-y_{i_\mathrm{target}}||^2,
    \label{lossfunc}
\end{equation}
\begin{equation*}
    \mathcal{L}_\mathrm{smoothness}=\lambda_\mathrm{smooth}\big(\frac{1}{|m|}\sum_{i=0}^{m}w_{i,j}(y_{\mathrm{pred}_i}-y_{\mathrm{pred}_j})^2\big),
\end{equation*}
with $\mathcal{L}_\mathrm{regression}$ being the standard mean squared loss for regression tasks, and $\mathcal{L}_\mathrm{smoothness}$ being the Laplacian based graph smoothness loss that encourages the vertices $i,j$  connected with higher weight $w_{i,j}$ to have similar values~\cite{GraphSignal}. $\lambda_\mathrm{smooth}$ is an arbitrary regularization parameter that allows the model to avoid overfitting by gradually increasing it during the training process. Another reason for including smoothness loss is that we want to incorporate that the higher orders of interaction account more towards the noise in fitness data than the lower orders. A custom loss function helps improve the generalization of the model. 
\section{\label{Results} Results}
We built a deep learning model using a graph convolution network with a multi-layer perceptron, and used wavelet transform for pre-processing of available data. We ran four experiments on different chemical activities for $419$ mutant sequences of \textit{N. Tabacum} enzyme out of $512$ all possible sequences. We use the $\mathit{db3}$ wavelet of Daubechie's' wavelet family as a wavelet basis for decomposition and denoising of the fitness signal using first-level DWT. The reconstructed denoised signal is passed as known labels for $419$ vertices, and the $9d$ binary feature vector for all $512$ vertices is passed to the GCN model. The first layer of the GCN model expands the $9$ dimensional input feature vector to $ 512$-dimensional hidden features, which is then passed to the second layer with a dropout rate of $0.6$ to be shrunken into a $ 256$-dimensional hidden feature vector. The MLP layer receives the hidden representation and outputs a unique fitness label for all vertices of the graph.

For the learning process, we introduce a training mask over all the vertices of the graph. We use $287$ randomly selected vertices at seed $40$ as training nodes, such that all training vertices contain fitness labels and are used to compute the loss $\mathcal{L}$ in Eq.~\ref{lossfunc}. The rest of the $132$ labeled vertices out of the total $419$ labeled vertices are kept as test vertices to validate the training later. $93$ unlabeled vertices and $132$ labeled test vertices are part of information aggregation but were not part of the training process. A standard Adam optimizer is used with learning rate and weight decay set at $5\times10^{-3}$ and $5\times10^{-4}$ respectively, to minimize the loss function by making iterative improvements to the model parameters. Lastly, the $\lambda_{\mathrm{smooth}}$, which is an arbitrary and dynamic scalar value, which means it changes during the training process is set to increase from $1.0$ to $10.0$, depending on $\min(10.0,\text{epoch}/100)$. This allows us to ensure that the model prioritizes regression over smoothness for initial epochs and then prioritizes smoothness. We shall now look at the results of training the model for $500$ epochs on different fitness labels with the same parameters.
\subsection{\label{Wavelet Denoising} Wavelet Denoising}
\begin{figure}[!ht]
    \centering
    \subfloat[\label{Denoised5EA}]{\includegraphics[width=\linewidth,height=0.2\linewidth]{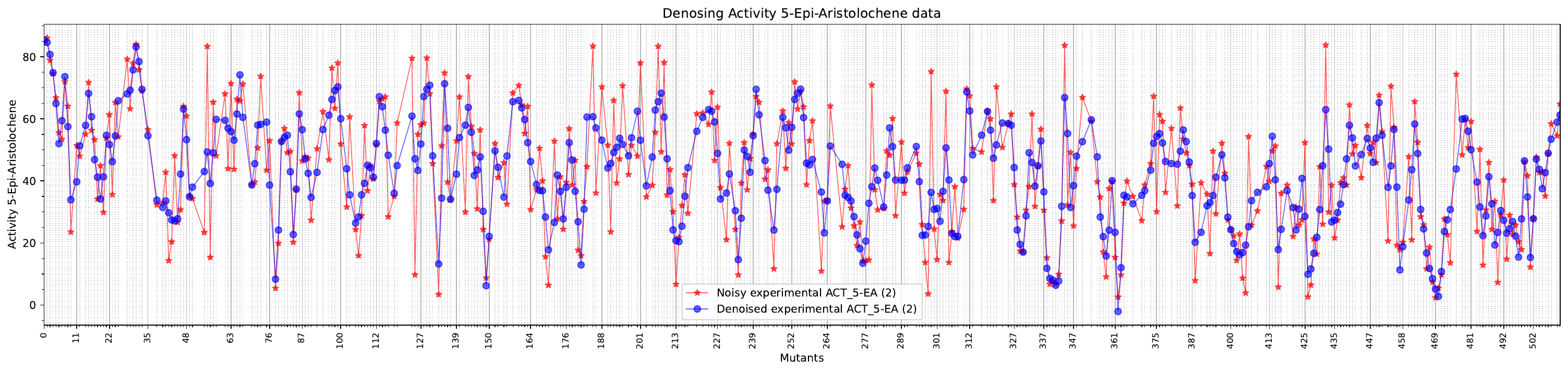}}
    \vspace{0.5mm}
    \subfloat[\label{Denoise4EE}]{\includegraphics[width=\linewidth,height=0.2\linewidth]{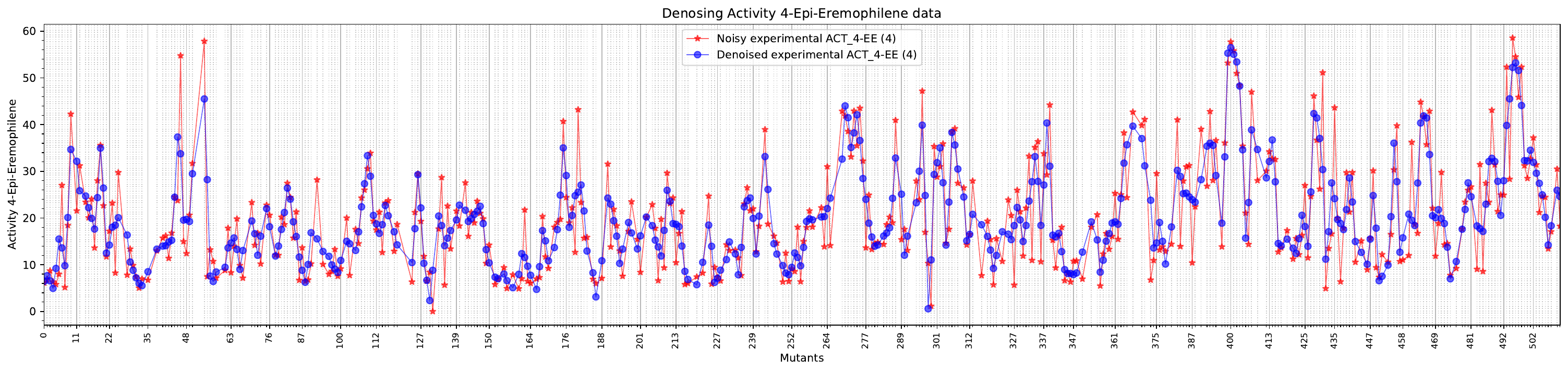}}
    \vspace{0.5mm}
    \subfloat[\label{DenoisedPSD}]{\includegraphics[width=\linewidth,height=0.2\linewidth]{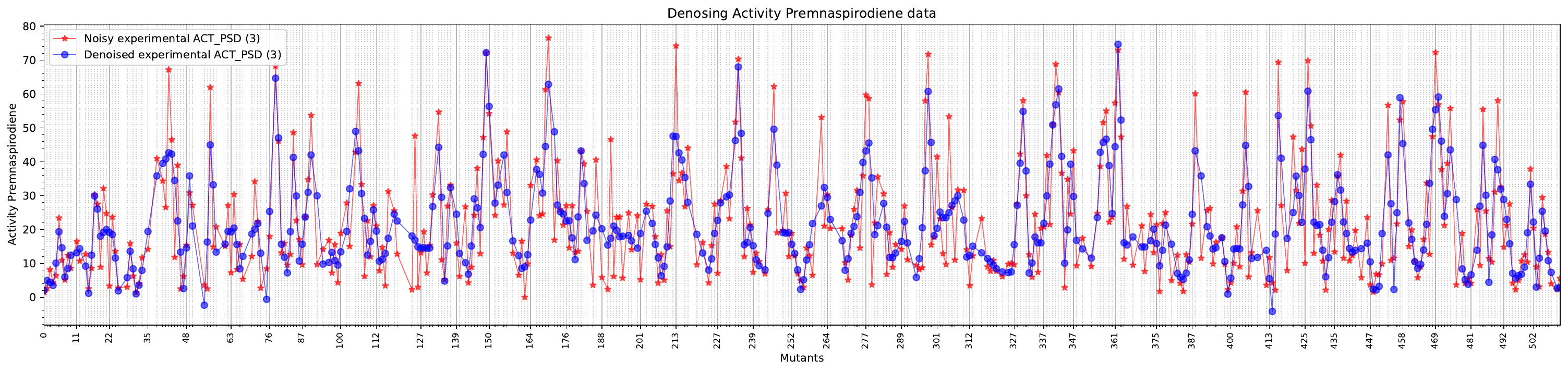}}
    \vspace{0.5mm}
    \subfloat[\label{DenoisedMP}]{\includegraphics[width=\linewidth,height=0.2\linewidth]{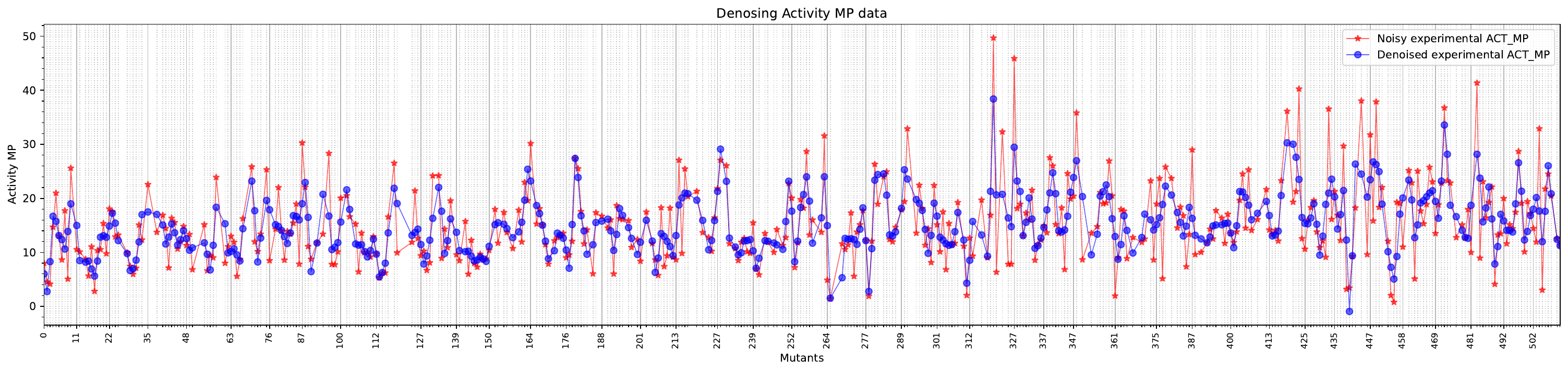}}
    \caption{A comparison of the wavelet denoised signal and the original signal for the production of various molecules by Tobacco 5-Epi-Aristolochlene Synthase (TEAS) as percentage of a gas chromatogram-mass spectrometry (GC-MS) data. Mutants are represented via numbers on the horizontal axis maps to a unique discrete fitness value on the vertical axis represented by a star or a circle. A connecting line makes it easier to follow the pattern it does not mean interpolation in between mutation is allowed. The missing sequences are shown via the gaps in the grid lines. The plot shows the percentage of total GC-MS data for the production of (a) 5-Epi-Aristolochene molecule, (b) 4-Epi-Eremophilene, (c) premnaspirodiene, and (d) minor products for each of the detected mutants.}
    \label{denoising}
\end{figure}

The fitness function constitutes a discrete signal because it defines a scalar-valued function on a discrete-structured domain that can be interpreted as the vertices of a Hamming graph. Lack of data restricts us from getting the exact understanding of the non-linear dynamics. As we have mentioned earlier in Sec.~\ref{Data denoising}, it is a challenging task to remove the global experimental noise in the data, which is fatal to the deep learning model and can hinder its learning capabilities. Therefore, we use the wavelet transform to minimize the variations in the fitness signal. A variation in the fitness signal or the ruggedness of the fitness space also indicates the presence of epistasis and can not be separated. 


Decomposition of the fitness signal can be interpreted as global and local signals. Global behavior of the signal is kept intact, and local variations are diminished via a threshold. Shrinking the small detail coefficients allows the sudden sharp local changes to vanish, revealing a much smoother underlying signal, as shown in Fig.~\ref{denoising}. We applied the method to four fitness signals namely, $5$-Epi-Aristolochene ($5$-EA) in Fig.~\ref{Denoised5EA}, $4$-Epi-Eremophilene ($4$-EE) in Fig.~\ref{Denoise4EE}, Premnaspirodiene (PSD) in Fig.~\ref{DenoisedPSD}, and Fig.~\ref{DenoisedMP} for the minor products. We see that the denoised signal is much smoother than the original signal, while the structure of the original signal is preserved, which shows that the inherent epistatic behavior of mutations is still intact. Similarity among the denoised and noisy experimental data, provides confidence that the sequence is truly functional and its fitness is due to local sequence-fitness relation rather than being some erratic behavior of individual sequence~\cite{minepi}.
It is worth noting that the smoother the signal is, meaning that the local variance is smaller, the better chances the GCN model has in terms of generalization. 
\subsection{\label{model performance} EHCube4P performance}
An Ablation study was conducted to test the role of three elements of the model: (1) Role of wavelet denoising, (2) Role of smoothness, (3) Role of MLP. Table.\ref{table one} summarizes our findings, tested on the $4$-Epi-Eremophilene activity, with a division of $287$ sequences as training mutants and $132$ sequences as test mutants. The finding suggests that wavelet denoising of the experimental data significantly amplifies the model's learnability. Introducing smoothing loss causes the Root Mean Square Error (RMSE) to decrease, which helps the model to predict close to accurate values for test mutants. Lastly, we show that the use of a multi-layer perceptron improves the model's capabilities to recognize the local fluctuations that help accurately predict unseen sequences during training. The plots for the study are shown in Supp.~Fig.~S2.

\begin{table}[ht]
    \centering
    \resizebox{\textwidth}{!}{
    \begin{tabular}{c||c||c||c}
        \toprule
        \textbf{Experiment} & \textbf{$\mathbf{R^2}$(test)} & \textbf{RMSE} & \textbf{Variations}\\\midrule
        Full Model (with Wavelet) & $\mathbf{0.66}$ & $\mathbf{5.85}$ & Complete model as manuscript \\
        \midrule
        Full Model (without Wavelet) & $0.38$ & $9.88$ &  without preprocessing of phenotype signal \\
        \midrule
        Full Model (without Smoothness loss)& $0.61$ & $6.19$ & Trained only on MSE \\
        \midrule
        Full Model (fixed $\lambda_\mathrm{smooth}$)& $0.62$ & $6.18$ & Trained with Eq.\ref{lossfunc} but fixed $\lambda_\mathrm{smooth}=1.0$ \\
        \midrule
        Full Model (dynamic but small $\lambda_\mathrm{smooth}$)& $0.62$ & $6.20$ & Trained with Eq.\ref{lossfunc} but varying $\lambda_\mathrm{smooth}\in[0.01,1.0]$ \\
        \midrule
        Full Model (dynamic but large $\lambda_\mathrm{smooth}$)& $0.66$ & $5.85$ & Trained with Eq.\ref{lossfunc} but varying $\lambda_\mathrm{smooth}\in[10.0,50.0]$ \\
        \midrule
        No MLP (with wavelet) & $0.54$ & $6.77$ & MLP layer removed\\
        \midrule
        No MLP (without wavelet) & $0.39$ & $9.83$ & No MLP and No wavelet preprocessing \\
        \bottomrule
        
    \end{tabular}}
    \caption{Ablation study for EHCube4P. All variations are tested with enzyme activity for the production of $4$-Epi-Eremophilene. The reported $\mathbf{R^2}$ for test set include $132$ mutant sequences.}
    \label{table one}
\end{table}

\begin{figure}[!ht]
    \centering
    \subfloat[]{\includegraphics[width=0.49\linewidth,height=0.3\linewidth]{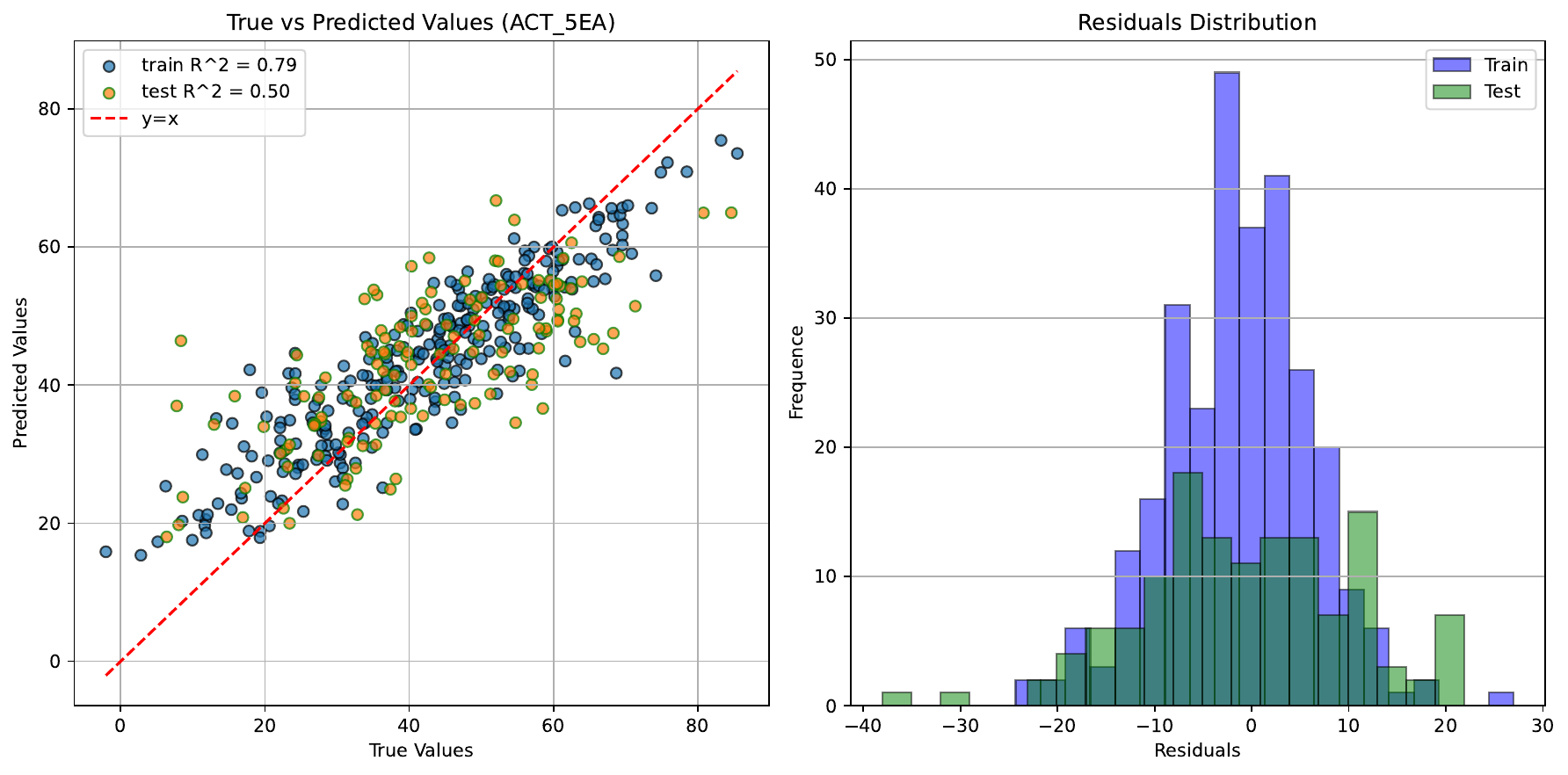}\label{Result5EA}}
    \hfill
    \subfloat[]{\includegraphics[width=0.49\linewidth,height=0.3\linewidth]{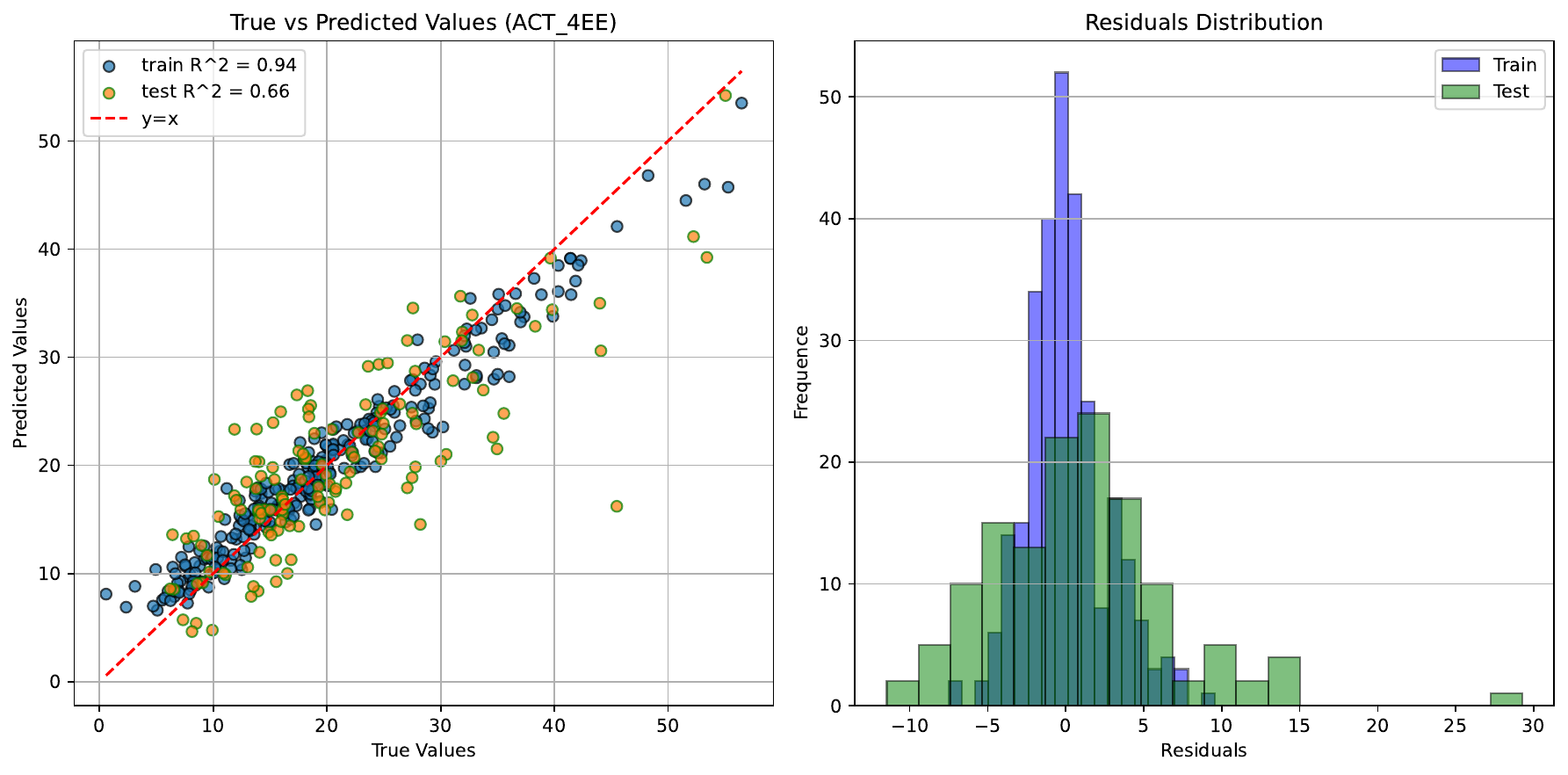}\label{Result4EE}}
    \vspace{1mm}
    \subfloat[]{\includegraphics[width=0.49\linewidth,height=0.3\linewidth]{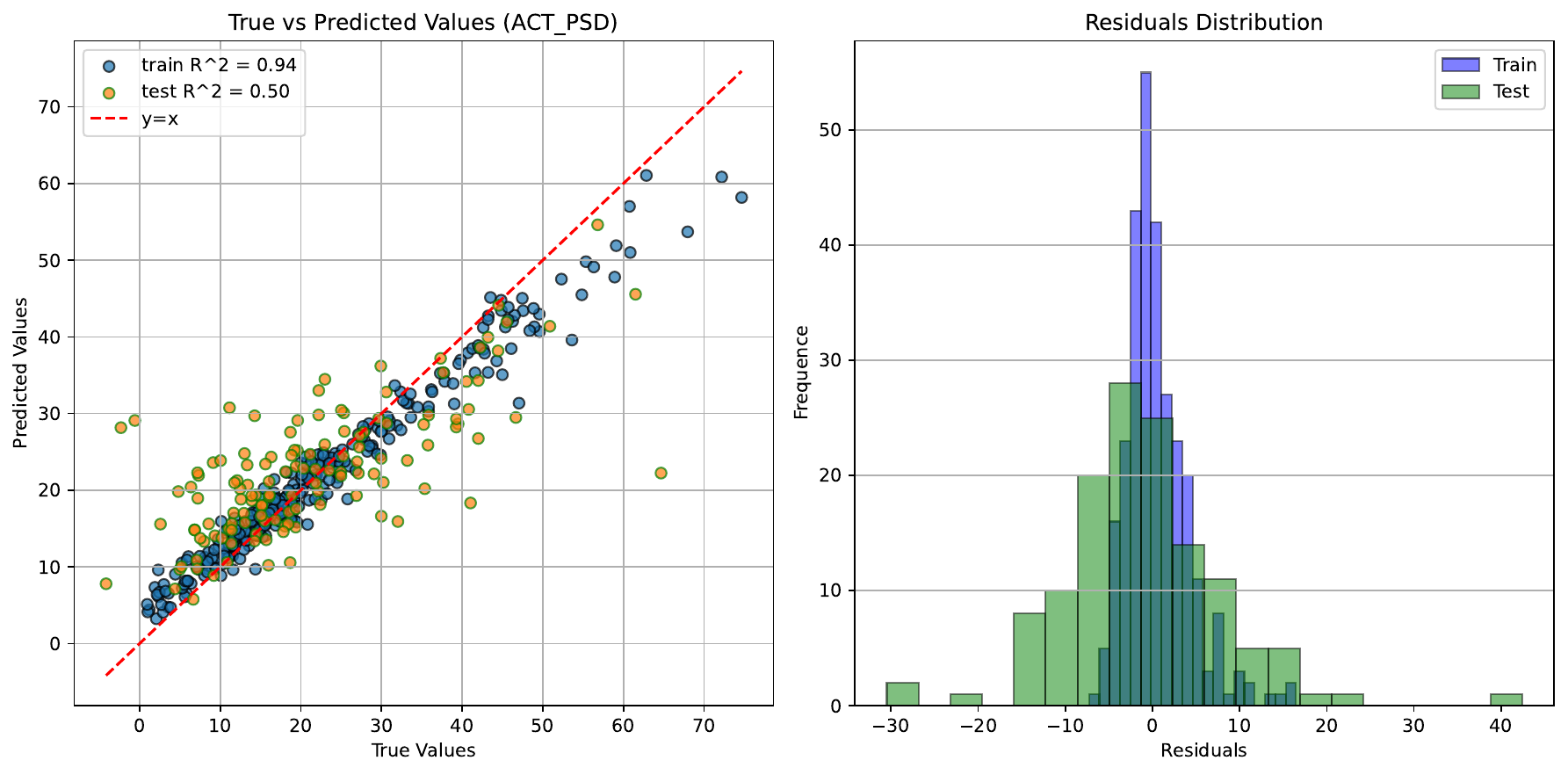}\label{ResultPSD}}
    \hfill
    \subfloat[]{\includegraphics[width=0.49\linewidth,height=0.3\linewidth]{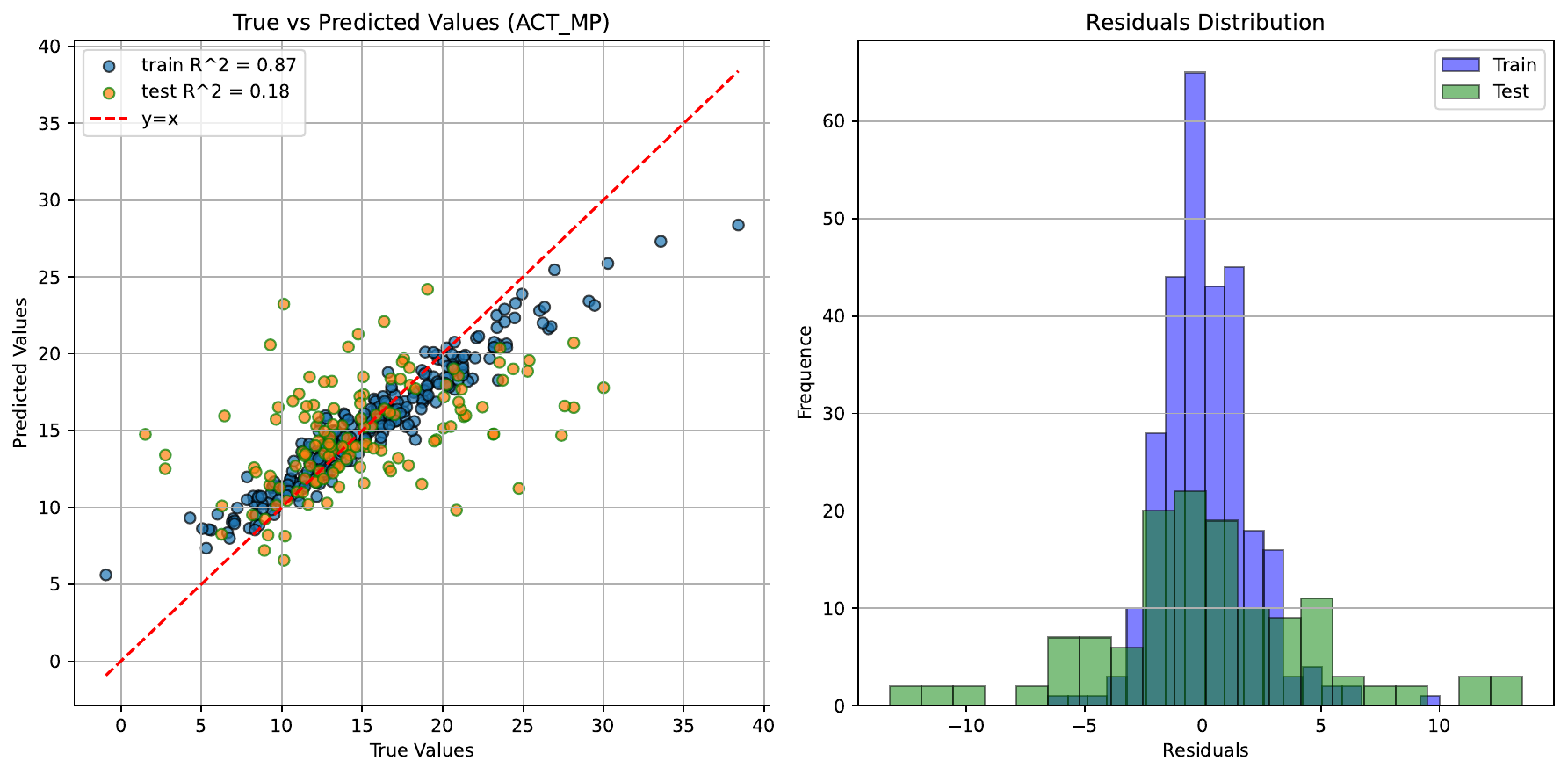}\label{ResultMP}}
    \caption{Model performance on four different enzyme activities while the model parameters are kept the same. The figure shows the correlation between the prediction of denoised experimental data and the residual distribution of the prediction for (a) $5$EA, (b) $4$EE, (c) PSD, and (4) MP.}
    \label{model stats}
\end{figure}
We ran four separate experiments, one for each fitness signal, with the best results from the ablation study. The results show great potential for GCN in predicting protein fitness estimation. The aggregation of information from adjacent vertices allows the features to capture beyond pairwise interactions. Two layers of aggregations allow the model to predict fitness values with confidence. We also see that the model's performance worsens on the test set as the variation in fitness values increases. We report that the GCN model is able to learn epistatic patterns from the dataset and use them to make sound predictions about the fitness values of a mutant sequence. We show that the best results were achieved for the enzyme activity for the production of $4$-Epi-Eremophilene, in Fig.~\ref{Result4EE}, with a training and testing $r^2$ of $0.94$ and $0.66$, respectively. In Fig.~\ref{Result5EA} and Fig.~\ref{ResultPSD}, we report that the model achieved an $r^2$ value of $0.79$ and $0.94$ on training sets for the production of $5$-Epi-Aristolochlene and premnaspirodiene, respectively. While on the test data, both sets performed similarly with an $r^2$ of $0.50$. This result can be explained by the ordering of the signal and the permutation invariance of the GCNs. The reported results show that the model can capture the interaction trends when mutating from TEAS to TEAS-M9 or vice versa. The model predicts the productivity of $4$-Epi-Eremophilene with greater confidence because it shares the mutation background from both TEAS (wildtype) and TEAS-M9 (fully-mutated variant).

Lastly, Fig.~\ref{ResultMP} reports an $r^2$ of $0.87$ and $0.18$ on the training and testing sets, respectively. The results reported in Fig.~\ref{ResultMP} are worse off than the results in Fig.~\ref{Result5EA}, Fig.~\ref{Result4EE}, and Fig.~\ref{ResultPSD} because the smoothness loss $\mathcal{L}_\mathrm{smoothness}$ used in the training process encourages the connected vertices to have similar fitness values. Thus, the productivity data of minor products have erratic local behaviors, suggesting most data are not entirely functional, and we can not conclude with certainty that they are due to epistasis or some individual behavior of sequences. Though we can see that this approach is beneficial for datasets where mutationally adjacent sequences follow the same trajectory, i.e. sign epistasis or strong local variations are not in abundance. Such variations cause the model to perform poorly when the fitness landscape is too rugged.

\section{\label{Conc}Conclusion}
In this study, we demonstrated an effective framework for protein fitness prediction by integrating wavelet-based signal denoising with graph convolutional networks (GCNs) and multi-layer perceptrons (MLPs). Through the application of the discrete wavelet transform with a Daubechies-3 basis, we successfully reduced experimental noise in the enzyme activity data, revealing smoother underlying fitness landscapes without sacrificing critical structural patterns associated with epistatic interactions. By training a GCN model on 419 labeled mutants of the Tobacco 5-Epi-Aristolochlene Synthase (TEAS) enzyme, we showed that wavelet-preprocessed signals substantially improved the model's ability to generalize. The two-layer GCN architecture, followed by an MLP, effectively aggregated information across the mutant space, capturing non-trivial mutation patterns beyond simple pairwise interactions.

Our results indicate that EHCube4P performance was closely tied to the smoothness of the fitness landscape: datasets exhibiting lower local variance, such as the 4-Epi-Eremophilene production data, achieved higher predictive accuracy on the test set ($r^2=0.66$). In contrast, datasets with higher levels of local fluctuations, such as Premnaspirodiene (PSD) production, posed greater challenges for generalization, as reflected by lower test set $r^2$ values. Furthermore, we observed that dynamic tuning of the smoothness regularization parameter $\lambda_\mathrm{smooth}$ was critical to balancing the bias-variance trade-off, particularly in fitness landscapes with varying degrees of ruggedness. Excessive smoothness penalization negatively impacted model performance in highly epistatic or rugged datasets, while appropriate calibration of $\lambda_\mathrm{smooth}$ helped recover predictive performance. Overall, this work highlights the importance of signal pre-processing and structural regularization in improving predictive models for biological fitness landscapes. Future extensions could include adaptive wavelet thresholding strategies and deeper GNN architectures capable of capturing higher-order epistatic effects in more complex mutational datasets.

\section*{Author Contributions}
MD contributed to conceptualizing, experimenting, analyzing, and writing the manuscript. FC contributed to conceptualization, data procurement, analysis, and writing. PC helped with analyzing and writing. CD helped with analyzing and writing, and JW contributed to the discussions and helped with the proofreading and structuring of the manuscript.
\section*{Funding Declaration}
MD is supported by a PhD grant from the Region Reunion and European Union (FEDER-FSE 2021/2027) 2023062, 345879, also by UWA Scholarship for International Research Fee~(SIRF). PEACCEL was supported through a research program partially co-funded by the European Union (UE) and Region Reunion (FEDER).
\section*{Conflict of Interest Statement}
Author FC is linked to the company Peaccel.

\section*{Data Availability Statement}
The datasets used and/or analyzed during the current study are available from the corresponding author on reasonable request. The code and the dataset will be released publicly upon acceptance. The code used in this study will be publicly available under the Creative Commons Attribution-NonCommercial-NoDerivatives (CC BY-NC-ND 4.0) license upon publication.

\bibliography{ref.bib}

\begin{thebibliography}{10}
\providecommand{\doi}[1]{\url{https://doi.org/#1}}
\bibcommenthead

\bibitem[\protect\citeauthoryear{Acevedo-Rocha et~al.}{2021}]{Acevedo-Rocha2021}
Acevedo-Rocha CG, Li A, D'Amore L, Hoebenreich S, Sanchis J, Lubrano P, et~al.
\newblock Pervasive cooperative mutational effects on multiple catalytic enzyme traits emerge via long-range conformational dynamics.
\newblock Nature Communications. 2021 Mar;12(1):1621.
\newblock \doi{10.1038/s41467-021-21833-w}.

\bibitem[\protect\citeauthoryear{Freschlin et~al.}{2022}]{CRF2022}
Freschlin CR, Fahlberg SA, Romero PA.
\newblock Machine learning to navigate fitness landscapes for protein engineering.
\newblock Current Opinion in Biotechnology. 2022;75:102713.
\newblock \doi{https://doi.org/10.1016/j.copbio.2022.102713}.

\bibitem[\protect\citeauthoryear{Zhou et~al.}{2024}]{DLandPE}
Zhou B, Tan Y, Hu Y, Zheng L, Zhong B, Hong L.
\newblock Protein engineering in the deep learning era.
\newblock mLife. 2024;3(4):477--491.
\newblock \doi{https://doi.org/10.1002/mlf2.12157}.
\newblock {\href{https://arxiv.org/abs/https://onlinelibrary.wiley.com/doi/pdf/10.1002/mlf2.12157}{{https://onlinelibrary.wiley.com/doi/pdf/10.1002/mlf2.12157}}}.

\bibitem[\protect\citeauthoryear{Cadet et~al.}{2022}]{Cadet2022}
Cadet XF, Gelly JC, van Noord A, Cadet F, Acevedo-Rocha CG.
\newblock In: Currin A, Swainston N, editors. Correction to: Learning Strategies in Protein Directed Evolution. New York, NY: Springer US; 2022. p. C1--C1.
\newblock Available from: \url{https://doi.org/10.1007/978-1-0716-2152-3_16}.

\bibitem[\protect\citeauthoryear{Li et~al.}{2025}]{Wenran2024}
Li W, Cadet XF, Medina-Ortiz D, Davari MD, Sowdhamini R, Damour C, et~al.: From thermodynamics to protein design: Diffusion models for biomolecule generation towards autonomous protein engineering.
\newblock Available from: \url{https://arxiv.org/abs/2501.02680}.

\bibitem[\protect\citeauthoryear{Wittmund et~al.}{2022}]{Wittmund2022}
Wittmund M, Cadet F, Davari MD.
\newblock Learning Epistasis and Residue Coevolution Patterns: Current Trends and Future Perspectives for Advancing Enzyme Engineering.
\newblock ACS Catalysis. 2022 Nov;12(22):14243--14263.
\newblock \doi{10.1021/acscatal.2c01426}.

\bibitem[\protect\citeauthoryear{Romero and Arnold}{2009}]{Romero2009}
Romero PA, Arnold FH.
\newblock Exploring protein fitness landscapes by directed evolution.
\newblock Nature Reviews Molecular Cell Biology. 2009 Dec;10(12):866--876.
\newblock \doi{10.1038/nrm2805}.

\bibitem[\protect\citeauthoryear{Chowdhury and Garai}{2017}]{MSACHOWDHURY2017419}
Chowdhury B, Garai G.
\newblock A review on multiple sequence alignment from the perspective of genetic algorithm.
\newblock Genomics. 2017;109(5):419--431.
\newblock \doi{https://doi.org/10.1016/j.ygeno.2017.06.007}.

\bibitem[\protect\citeauthoryear{McCandlish}{2011}]{VisualingFitnessSpace}
McCandlish DM.
\newblock VISUALIZING FITNESS LANDSCAPES.
\newblock Evolution. 2011 06;65(6):1544--1558.
\newblock \doi{10.1111/j.1558-5646.2011.01236.x}.
\newblock {\href{https://arxiv.org/abs/https://academic.oup.com/evolut/article-pdf/65/6/1544/47949872/evolut1544.pdf}{{https://academic.oup.com/evolut/article-pdf/65/6/1544/47949872/evolut1544.pdf}}}.

\bibitem[\protect\citeauthoryear{Poelwijk et~al.}{2019}]{Poelwijk2019}
Poelwijk FJ, Socolich M, Ranganathan R.
\newblock Learning the pattern of epistasis linking genotype and phenotype in a protein.
\newblock Nature Communications. 2019 Sep;10(1):4213.
\newblock \doi{10.1038/s41467-019-12130-8}.

\bibitem[\protect\citeauthoryear{Cadet et~al.}{2018}]{FCadet2018}
Cadet F, Fontaine N, Li G, Sanchis J, Ng~Fuk~Chong M, Pandjaitan R, et~al.
\newblock A machine learning approach for reliable prediction of amino acid interactions and its application in the directed evolution of enantioselective enzymes.
\newblock Scientific Reports. 2018 Nov;8(1):16757.
\newblock \doi{10.1038/s41598-018-35033-y}.

\bibitem[\protect\citeauthoryear{Levy et~al.}{2017}]{LEVY201755}
Levy RM, Haldane A, Flynn WF.
\newblock Potts Hamiltonian models of protein co-variation, free energy landscapes, and evolutionary fitness.
\newblock Current Opinion in Structural Biology. 2017;43:55--62.
\newblock Theory and simulation • Macromolecular assemblies. \doi{https://doi.org/10.1016/j.sbi.2016.11.004}.

\bibitem[\protect\citeauthoryear{Biswas et~al.}{2021}]{Biswas2021}
Biswas S, Khimulya G, Alley EC, Esvelt KM, Church GM.
\newblock Low-N protein engineering with data-efficient deep learning.
\newblock Nature Methods. 2021 Apr;18(4):389--396.
\newblock \doi{10.1038/s41592-021-01100-y}.

\bibitem[\protect\citeauthoryear{Hsu et~al.}{2022}]{Hsu2022}
Hsu C, Nisonoff H, Fannjiang C, Listgarten J.
\newblock Learning protein fitness models from evolutionary and assay-labeled data.
\newblock Nature Biotechnology. 2022 Jul;40(7):1114--1122.
\newblock \doi{10.1038/s41587-021-01146-5}.

\bibitem[\protect\citeauthoryear{Zhou et~al.}{2022}]{HOE&PP}
Zhou J, Wong MS, Chen WC, Krainer AR, Kinney JB, McCandlish DM.
\newblock Higher-order epistasis and phenotypic prediction.
\newblock Proceedings of the National Academy of Sciences. 2022;119(39):e2204233119.
\newblock \doi{10.1073/pnas.2204233119}.
\newblock {\href{https://arxiv.org/abs/https://www.pnas.org/doi/pdf/10.1073/pnas.2204233119}{{https://www.pnas.org/doi/pdf/10.1073/pnas.2204233119}}}.

\bibitem[\protect\citeauthoryear{Sanchez-Lengeling et~al.}{2021}]{sanchez-lengeling2021a}
Sanchez-Lengeling B, Reif E, Pearce A, Wiltschko AB.
\newblock A Gentle Introduction to Graph Neural Networks.
\newblock Distill. 2021;Https://distill.pub/2021/gnn-intro. \doi{10.23915/distill.00033}.

\bibitem[\protect\citeauthoryear{Zhang et~al.}{2021}]{GNN_IN_BI}
Zhang XM, Liang L, Liu L, Tang MJ.
\newblock Graph Neural Networks and Their Current Applications in Bioinformatics.
\newblock Frontiers in Genetics. 2021;Volume 12 - 2021.
\newblock \doi{10.3389/fgene.2021.690049}.

\bibitem[\protect\citeauthoryear{Wang et~al.}{2023}]{WANG2023107952}
Wang S, Tang H, Shan P, Wu Z, Zuo L.
\newblock ProS-GNN: Predicting effects of mutations on protein stability using graph neural networks.
\newblock Computational Biology and Chemistry. 2023;107:107952.
\newblock \doi{https://doi.org/10.1016/j.compbiolchem.2023.107952}.

\bibitem[\protect\citeauthoryear{O'Maille et~al.}{2008}]{O'Maille2008}
O'Maille PE, Malone A, Dellas N, Andes~Hess B, Smentek L, Sheehan I, et~al.
\newblock Quantitative exploration of the catalytic landscape separating divergent plant sesquiterpene synthases.
\newblock Nature Chemical Biology. 2008 Oct;4(10):617--623.
\newblock \doi{10.1038/nchembio.113}.

\bibitem[\protect\citeauthoryear{Koo et~al.}{2016}]{Koo2016}
Koo HJ, Vickery CR, Xu Y, Louie GV, O'Maille PE, Bowman M, et~al.
\newblock Biosynthetic potential of sesquiterpene synthases: product profiles of Egyptian Henbane premnaspirodiene synthase and related mutants.
\newblock The Journal of Antibiotics. 2016 Jul;69(7):524--533.
\newblock \doi{10.1038/ja.2016.68}.

\bibitem[\protect\citeauthoryear{Poelwijk}{2019}]{Poelwijkbook}
Poelwijk FJ.
\newblock In: Sikosek T, editor. Context-Dependent Mutation Effects in Proteins. New York, NY: Springer New York; 2019. p. 123--134.
\newblock Available from: \url{https://doi.org/10.1007/978-1-4939-8736-8_7}.

\bibitem[\protect\citeauthoryear{Schneider and Farge}{2006}]{SCHNEIDER2006426}
Schneider K, Farge M.
\newblock Wavelets: Mathematical Theory.
\newblock In: Françoise JP, Naber GL, Tsun TS, editors. Encyclopedia of Mathematical Physics. Oxford: Academic Press; 2006. p. 426--438.
\newblock Available from: \url{https://www.sciencedirect.com/science/article/pii/B012512666200153X}.

\bibitem[\protect\citeauthoryear{Mallat}{1999}]{mallat1999wavelet}
Mallat S.
\newblock A wavelet tour of signal processing.
\newblock Elsevier; 1999.

\bibitem[\protect\citeauthoryear{Guo et~al.}{2022}]{WTR}
Guo T, Zhang T, Lim E, López-Benítez M, Ma F, Yu L.
\newblock A Review of Wavelet Analysis and Its Applications: Challenges and Opportunities.
\newblock IEEE Access. 2022;10:58869--58903.
\newblock \doi{10.1109/ACCESS.2022.3179517}.

\bibitem[\protect\citeauthoryear{Moulin}{2005}]{MOULIN2005347}
Moulin P.
\newblock 4.2 - Multiscale Image Decompositions and Wavelets.
\newblock In: BOVIK A, editor. Handbook of Image and Video Processing (Second Edition). second edition ed. Communications, Networking and Multimedia. Burlington: Academic Press; 2005. p. 347--359.
\newblock Available from: \url{https://www.sciencedirect.com/science/article/pii/B978012119792650084X}.

\bibitem[\protect\citeauthoryear{He et~al.}{2015}]{WT_UT}
He C, Xing J, Li J, Yang Q, Wang R.
\newblock A New Wavelet Threshold Determination Method Considering Interscale Correlation in Signal Denoising.
\newblock Mathematical Problems in Engineering. 2015;2015(1):280251.
\newblock \doi{https://doi.org/10.1155/2015/280251}.
\newblock {\href{https://arxiv.org/abs/https://onlinelibrary.wiley.com/doi/pdf/10.1155/2015/280251}{{https://onlinelibrary.wiley.com/doi/pdf/10.1155/2015/280251}}}.

\bibitem[\protect\citeauthoryear{Donoho and Johnstone}{1994}]{donoho1994ideal}
Donoho DL, Johnstone IM.
\newblock Ideal spatial adaptation by wavelet shrinkage.
\newblock biometrika. 1994;81(3):425--455.

\bibitem[\protect\citeauthoryear{Chatterjee and Walker}{2017}]{Chatterjee2017-up}
Chatterjee N, Walker GC.
\newblock Mechanisms of {DNA} damage, repair, and mutagenesis.
\newblock Environ Mol Mutagen. 2017 May;58(5):235--263.

\bibitem[\protect\citeauthoryear{Brown}{2002}]{Brown2002-uv}
Brown TA.
\newblock Mutation, repair and recombination.
\newblock New York, NY: Wiley-Liss; 2002.

\bibitem[\protect\citeauthoryear{Poelwijk et~al.}{2016}]{Poelwijk2016}
Poelwijk FJ, Krishna V, Ranganathan R.
\newblock The Context-Dependence of Mutations: A Linkage of Formalisms.
\newblock PLOS Computational Biology. 2016 06;12(6):1--19.
\newblock \doi{10.1371/journal.pcbi.1004771}.

\bibitem[\protect\citeauthoryear{Gori et~al.}{2005}]{gori2005new}
Gori M, Monfardini G, Scarselli F.
\newblock A new model for learning in graph domains.
\newblock In: Proceedings. 2005 IEEE international joint conference on neural networks, 2005.. vol.~2. IEEE; 2005. p. 729--734.

\bibitem[\protect\citeauthoryear{Kipf and Welling}{2017}]{kipf2017semisupervised}
Kipf TN, Welling M.
\newblock Semi-Supervised Classification with Graph Convolutional Networks.
\newblock In: International Conference on Learning Representations; 2017. Available from: \url{https://openreview.net/forum?id=SJU4ayYgl}.

\bibitem[\protect\citeauthoryear{Ma et~al.}{2019}]{ma2019flexiblegenerativeframeworkgra}
Ma J, Tang W, Zhu J, Mei Q.
\newblock A Flexible Generative Framework for Graph-based Semi-supervised Learning.
\newblock In: Wallach H, Larochelle H, Beygelzimer A, d\textquotesingle Alch\'{e}-Buc F, Fox E, Garnett R, editors. Advances in Neural Information Processing Systems. vol.~32. Curran Associates, Inc.; 2019. Available from: \url{https://proceedings.neurips.cc/paper_files/paper/2019/file/e0ab531ec312161511493b002f9be2ee-Paper.pdf}.

\bibitem[\protect\citeauthoryear{Shuman et~al.}{2013}]{GraphSignal}
Shuman DI, Narang SK, Frossard P, Ortega A, Vandergheynst P.
\newblock The emerging field of signal processing on graphs: Extending high-dimensional data analysis to networks and other irregular domains.
\newblock IEEE Signal Processing Magazine. 2013;30(3):83--98.
\newblock \doi{10.1109/MSP.2012.2235192}.

\bibitem[\protect\citeauthoryear{Zhou and McCandlish}{2020}]{minepi}
Zhou J, McCandlish DM.
\newblock Minimum epistasis interpolation for sequence-function relationships.
\newblock Nature Communications. 2020 Apr;11(1):1782.
\newblock \doi{10.1038/s41467-020-15512-5}.

\end{thebibliography}

\end{document}